\documentclass[journal,10pt]{IEEEtran}
\usepackage{epsfig}
\usepackage{amssymb}
\usepackage{latexsym,graphicx}
\usepackage{longtable}
\usepackage{color}
\usepackage{hyperref}
\usepackage{array}
\usepackage{mathrsfs}
\usepackage{amsmath}
\usepackage{ifthen}
\usepackage{array,color}
\usepackage{tabularx}
\usepackage[
inline
]{trackchanges}

\def\BibTeX{{\rm B\kern-.05em{\sc i\kern-.025em b}\kern-.08em
    T\kern-.1667em\lower.7ex\hbox{E}\kern-.125emX}}

\definecolor{navy}{RGB}{0,0,128}

 \makeatletter
 \let\NAT@parse\undefined
 \makeatother

\begin{document}
%
\title{Toward AIML Enabled WiFi Beamforming CSI Feedback Compression: An Overview of IEEE 802.11 Standardization}
%
\author{Ziming He, {\em Senior Member, IEEE}
\thanks{The author is with Samsung Cambridge Solution Centre, System LSI, Samsung Electronics,
One Cambridge Sq, Milton Ave, Cambridge CB4 0AE
e-mail: ziming.he@samsung.com.}}
\markboth{} {Shell IEEE \MakeLowercase{\textit{et al.}}: Bare Demo of
IEEEtran.cls for Journals}\maketitle

\begin{abstract}
Transmit beamforming is one of the key techniques used in the existing IEEE 802.11 WiFi standards and future generations such as 11be and 11bn, a.k.a., ultra high reliability (UHR).
The paper gives an overview of the current standardization activities regarding the artificial intelligence and machine learning (AIML) enabled beamforming channel state information (CSI)
feedback compression technique, defined by the 802.11 AIML topic interest group (TIG).
Two key challenges the AIML TIG is going to tackle in the future beamforming standards and four defined key performance indicators (KPIs) for the AIML enabled schemes are discussed in the paper. 
The two challenges are the CSI feedback overhead and the compression complexity, and the four KPIs are feedback overhead, AIML model sharing overhead, packet error rate and complexity.
Moreover, the paper presents a couple of AIML enabled compression schemes accepted by the TIG, such as the K-means and autoencoder based schemes,
and uses simulated and analyzed data to explain how these schemes are designed according to the KPIs. Finally, future research directions are indicated for encouraging more researchers and engineers to contribute to this technique and the standardization of the next generation WiFi beamforming.
\end{abstract}


\section{Introduction}\label{intro_sec}
Transmit beamforming (BF) is one of the key techniques used in the the existing WiFi standards since IEEE 802.11n \cite{wifi6} and it will continue to appear in the next generations, such as 11be and 11bn 
(i.e., WiFi 7 and 8). 11be is currently under standardization in its task group (TG) to develop the amendment, and beamforming has already been considered as a feature. 
The standardization for 11bn is still at its early stage, known as the ultra high reliability (UHR) study group (SG). The beamforming enhancement in UHR SG has recently been discussed in \cite{ZLinUHR2023}.

On the other hand, a topic interest group (TIG) is formed in July 2022 to discuss possible support of artificial intelligence and machine learning (AIML) by 802.11 standards.
The AIML TIG targets to describe use cases for AIML applicability in 802.11 systems and investigate the technical feasibility of features enabling the support of AIML \cite{XWangUHR2023}. 
Some of the use cases defined by AIML TIG are closely related to what are being considered for UHR.
Using AIML to enhance WiFi beamforming is currently the most popular use case that has been defined by the AIML TIG, key contributors are Samsung, Huawei and InterDigital \cite{XWangAIMLRprt2023}.
The AIML model sharing is another important use case to consider how to share AIML models in WiFi systems, the use case is not only related to the beamforming use case, but also related to
the other use cases that have already been defined by the AIML TIG, such as channel access and roaming enhancements \cite{XWangAIMLRprt2023}. 
The AIML TIG targets to conclude in September or November 2023 and has a chance to impact UHR SG or future WiFi standards.

\begin{figure*}[ht]
\begin{center}
\includegraphics[scale=0.10,width=\linewidth,trim=0cm 8cm 0cm 0cm,clip]{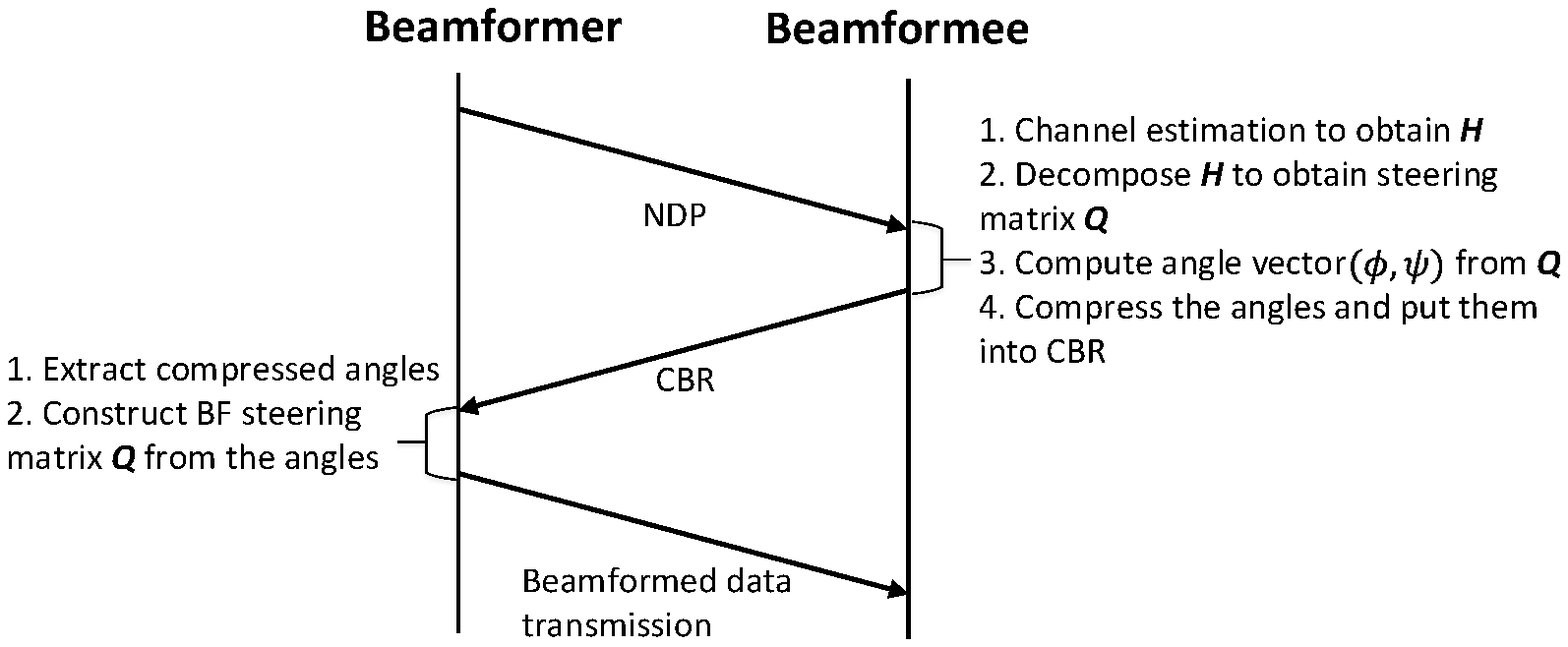}
\end{center}
\caption{A simplified WiFi beamforming procedure in IEEE 802.11ax}
\label{leg_mthd}
\end{figure*}

Beamforming applies different phase shifts to different transmit antennas, respectively, according to a steering matrix. 
With beamforming, a transmission is steered to a particular direction, in order to achieve an improved effective signal-to-noise ratio (SNR) at the receiver. 
The device that applies the steering matrix at its transmitter is referred to as beamformer, and the device that receives the beamformed transmission is referred to as beamformee.
A simplified WiFi beamforming procedure in 11ax is depicted in Fig. \ref{leg_mthd}:
\begin{itemize}
\item Firstly, the beamformer transmits a null data packet (NDP) for the beamformee to estimate the wireless channel and calculate the steering matrix $\mathbf{Q}$ by decomposing the channel estimate matrix
$\mathbf{H}$. $\mathbf{Q}$ can be seen as the channel state information (CSI) \cite{ZLinKeans2022}.
The NDP contains preamble but does not carry any user-specific data, it is used to enable preamble-based channel sounding.
\item Secondly, the beamformee sends the compressed angles ($\phi$ and $\psi$) computed from $\mathbf{Q}$, to the beamformer in a compressed beamforming report (CBR). 
The process to generate the information bits representing $\mathbf{Q}$ (e.g., the bits representing the compressed angles) is known as CSI feedback compression.
\item Finally, the beamformer receives the angles from CBR to reconstruct $\mathbf{Q}$, and then used it for beamformed data transmission. 
\end{itemize}
The technique is based on the assumptions that the wireless propagation channel is reciprocity and the procedure takes place in a short enough time period so that the time-variant channel does not change too much.  If the channel experienced by the beamformed data transmission has large enough difference between the channel experienced by the NPD, the effective SNR reduces and the packet error rate (PER) of beamforming beamformed transmission increases. This is because the applied steering matrix does not match the one derived from the channel experienced by the beamformed transmission.

In WiFi systems, either an access point (AP) or a non-AP station (e.g., a mobile phone) can be a beamformer. 
It is more common for the AP to serve as the beamformer since it is often equipped with multiple antennas. 
The maximum defined number of transmit antennas at AP is $4$ in 11n, $8$ in 11ax and expected to be $16$ in 11be \cite{ZLinKeans2022}. 
There is a trend that the number of transmit antennas will be further increased in 11bn and future generations, and this brings two new challenges
for the design of next generation WiFi beamforming systems:
\begin{itemize}
\item {\bf Challenge I}: As the number of antennas increases, the number of information bits required to represent the compressed angles in CBR increases, i.e., CSI feedback overhead increases. 
This creates more communication overhead and may eventually reduce the system goodput \cite{ZLinKeans2022}. The goodput is defined as the number of user data bits successfully delivered from the beamformer to the beamformee during a beamforming procedure, e.g., the one described in Fig. \ref{leg_mthd}. Note that NDP and CBR do not carry user data and are considered as overhead.
\item {\bf Challenge II}: As the number of antennas increases, the computational complexity of CSI feedback compression increases significantly.
A higher complexity compression scheme causes larger CSI feedback processing delay and consumes more power at non-AP stations \cite{ZHeAE2023}. 
A large feedback delay may then transfer to large channel changes in the beamformed data transmission and lead to PER performance degradation.
\end{itemize}
Accordingly, the beamforming use case defined by AIML TIG are focusing on the CSI feedback compression, its objectives are to leverage AIML to improve the system goodput via reducing the CSI feedback overhead or/and reduce the computational complexity of CSI feedback compression without degrading the goodput performance \cite{XWangAIMLRprt2023}.

The rest of the paper is organized as follows: Sec. \ref{leg} introduces the legacy CSI feedback compression scheme in the existing WiFi standards, 
then followed by various AIML enabled compression schemes with performance results in Sec. \ref{aiml_sec}. 
Finally, Sec. \ref{future_sec} summarizes future research directions which have not been addressed by the AIML TIG, and Sec. \ref{concl_sec} concludes.

\section{Legacy CSI compression scheme in the existing standards}\label{leg}
The purpose of beamforming CSI compression is to deliver the steering matrix from the beamformee to the beamformer with acceptable precision and feedback overhead.
For a given $\mathbf{Q}$, a CSI compression scheme with higher overhead usually offers better precision in steering matrix delivery, but may or may not lead to higher goodput since the overhead need to be taken into account. In the existing WiFi standards, the logic of designing a good CSI compression scheme in terms of maximizing the goodput is to find 
the best trade-off between the CSI precision and the feedback overhead.

The legacy CSI compression scheme (e.g., in 11ax \cite{wifi6}) is to firstly compute angles $\phi$ and $\psi$ from the steering matrix $\mathbf{Q}$ using Givens rotation \cite{ZLinKeans2022}, 
then quantize the obtained angles to integers that can be represented by bits. The CSI compression need to be performed for all selected feedback subcarriers, each of which has 
a corresponding $\mathbf{Q}$ with size $N_r\times N_c$. $N_r$ and $N_c$ are the number of rows and columns of the matrix, respectively. $N_r$ is equal to the number of antennas at beamformer, 
and $N_c$ is equal to the number of spatial steams used in the beamformed data transmission. The number of computed angles per $\mathbf{Q}$ are the same for $\phi$ and $\psi$, and it is related to 
$N_r$ and $N_c$. The number of bits used to quantize a $\psi$ and $\phi$ are defined as $N_b$ and $(N_b+2)$, respectively, since the precision requirement for $\phi$ is higher.
The legacy scheme defines a coarse and fine compression modes with $N_b$ equals to $2$ and $4$, respectively. The larger $N_b$ provides a higher precision for the re-constructed $\mathbf{Q}$ at beamformer, 
and thus offer a higher effective SNR at beamformee and a better PER performance; the smaller $N_b$ provides a less CSI feedback overhead but its PER performance is worse. 
The objective of $N_b$ selection is to find a better trade-off between the CSI precision and the feedback overhead.

\section{AIML enabled CSI feedback compression schemes}\label{aiml_sec}
\subsection{General description}\label{general_sec}
The AIML enabled CSI compression schemes accepted by the AIML TIG can be divided into two categories, the K-means based schemes proposed in \cite{ZLinKeans2022}, \cite{ZHeKmeans2023} and \cite{JeonDualCSI2023}, and the autoencoder based schemes proposed in \cite{ZHeAE2023}, \cite{SangdehAE2020} and \cite{GuoAE2023}. These accepted schemes have been summarized in the AIML TIG technical report in \cite{XWangAIMLRprt2023}.

\begin{figure*}[ht]
\begin{center}
\includegraphics[scale=0.10,width=\linewidth,trim=0cm 0.5cm 0cm 0.5cm,clip]{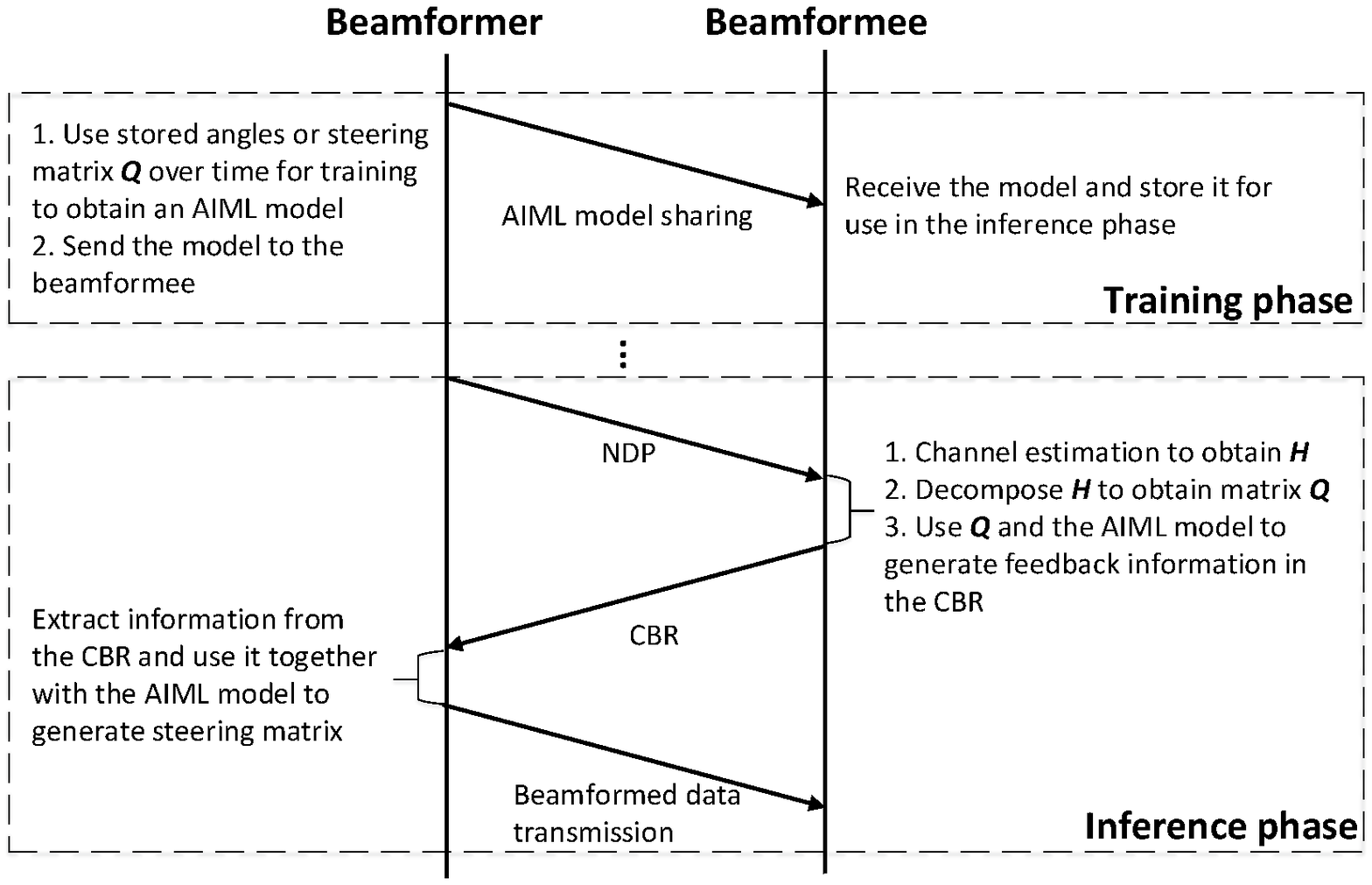}
\end{center}
\caption{A general description of beamforming procedure with AIML enable CSI compression}
\label{aiml_mthd}
\end{figure*}

The general procedure of the AIML compression schemes is depicted in Fig. \ref{aiml_mthd}, and it contains a training and an inference phase.
The training phase is to use the CSI data collected from the legacy beamforming procedure (defined by the previous standards, e.g., in Fig. \ref{leg_mthd}) 
to perform training and obtain an AIML model. After carrying out the legacy beamforming procedure multiple times, the beamformer can collect multiple angle vectors and steering matrices at various time instants, respectively. The CSI training data can be either the received angles or the re-constructed $\mathbf{Q}$, depending on a specific scheme.
The basic design requirement of new AIML enabled CSI compression standard is to provide backward compatibility to support the legacy beamforming procedure. 
Thus, utilizing the legacy beamforming procedure for training data collection is a reasonable solution to avoid additional signaling overhead introduced by the AIML schemes \cite{ZLinCSIUseCase2022}.
The beamformer is usually responsible for the training to offload the high computation burden to the AP side.
The trained model is sent to the beamformee, and then stored at the beamformee for use in the inference phase. This is known as the AIML model sharing.
Similar to the legacy procedure in Fig. \ref{leg_mthd}, the inference phase of the AIML schemes also includes NDP, CBR and beamformed data transmission, 
the difference is that the AIML model need to be utilized to generate the CBR and re-constructed steering matrix.

\subsection{K-means based schemes}\label{kmeans_sec}
K-means is an unsupervised learning approach to cluster vectors based on conventional machine learning techniques. 
If there are $N_v$ input vectors (each has length $M$) that used for training, the output from the K-means algorithm is $N_k$ vectors (each also has length $M$) presenting the centroids of $N_k$ clusters, respectively. Each cluster includes one or a couple of input vectors, and thus $N_k$ is usually less than $N_v$. The $N_k$ vectors (or centroids) are the training output and used for inference. During inference, the $N_k$ centroids is compared with an input vector to identify the most similar one, e.g., the smallest Euclidean distance between the centroids and the input \cite{ZLinKeans2022}. 
The index (from $1$ to $N_k$) of the found centroid is the output from the inference.

There are three basic K-means based CSI compression schemes in \cite{ZLinKeans2022} and \cite{ZHeKmeans2023} (or \cite{HeBFPatent1}) accepted by the AIML TIG. 
The difference in the training phase comes from the content of the vectors/centroids:
{\bf 1) Scheme in \cite{ZLinKeans2022}}: A training vector containing all angles $\phi$ and $\psi$ computed from a steering matrix $\mathbf{Q}$. 
A centroid represents either a vector containing both $\phi$ and $\psi$ after clustering.
{\bf 2) First scheme in \cite{ZHeKmeans2023}}: A training vector containing either angles $\phi$ or $\psi$ computed from a steering matrix $\mathbf{Q}$.
It is difficult to describe what a centroid represents after clustering, since the training input is a mix of $\phi$ or $\psi$ vectors.
{\bf 3) Second scheme in \cite{ZHeKmeans2023}}: A training vector containing of all real and imaginary values in a steering matrix $\mathbf{Q}$. Note that the elements in the last row of $\mathbf{Q}$ are real values. A centroid represents a vector containing $\mathbf{Q}$ after clustering.

The number of training vectors used in the scheme in \cite{ZLinKeans2022} and the second scheme in \cite{ZHeKmeans2023} are the same. Each training vector
comes from a $\mathbf{Q}$ at a feedback subcarrier and a NDP sounding time instant. 
The number of training samples (or vectors) used in the first scheme in \cite{ZHeKmeans2023} is double of that in the other two schemes, 
since there are two vectors coming from a feedback subcarrier and a NDP sounding, one representing $\phi$ and other representing $\psi$.
The output centroids from training is referred to as the codebook representing the trained AIML model, and a centroid of a cluster is referred to as a codeword.

The codebook needs to be delivered from the beamformer to the beamformee, i.e., the AIML model sharing step in Fig. \ref{aiml_mthd}. 
A simple codebook compression scheme is proposed in \cite{ZHeKmeans2023} for the three schemes by leveraging the existing standards:
{\bf 1) Scheme in \cite{ZLinKeans2022}}: Quantize the codebook using the existing method described in Sec. \ref{general_sec}, i.e., use $N_b$ bits to quantize $\psi$ and $(N_b+2)$ bits to quantize $\phi$.
{\bf 2) First scheme in \cite{ZHeKmeans2023}}: Quantize the codebook using $(N_b+2)$ bits. This is because it is not known whether a centroid element represents $\phi$ or $\psi$, 
it is better to use a higher precision.
{\bf 3) Second scheme in \cite{ZHeKmeans2023}}: Compute angles $\psi$ and $\phi$ from all codewords in the codebook using Givens rotation and then quantize the angles using the method described in {\bf 1)}.

The key difference in the inference phase is the processing step 3 at beamformee in Fig. \ref{aiml_mthd}.
The step are summarized below for the three schemes:
{\bf 1) Scheme in \cite{ZLinKeans2022}}: Compute $\phi$ and $\psi$ angles from $\mathbf{Q}$ and concatenate them to obtain a vector, and find the index of the most similar codeword with it in the codebook.
{\bf 2) First scheme in \cite{ZHeKmeans2023}}: Compute $\phi$ and $\psi$ from $\mathbf{Q}$ to get two vectors, respectively. 
Then, find the two indices of the most similar two codewords with the two vectors in the codebook, respectively. Note that the same codebook is used when searching the two indices.
{\bf 3) Second scheme in \cite{ZHeKmeans2023}}: Find the index of the most similar codeword with $\mathbf{Q}$ in the codebook. 

One way to find the most similar codeword is to find the smallest Euclidean distance between a received vector and all the codewords in the codebook.
The above step is repeated for all feedback subcarriers and all the found indices are then put into the CBR. 
The required number of feedback bits for a codeword index (i.e, $N_{bf}$) determines a $N_k$.
For example, when $N_{bf}=10$ bits, $N_k=1024$.

Apart from the three basic K-means schemes described above, a hybrid legacy and K-means scheme is proposed in \cite{JeonDualCSI2023}. 
The key idea is to decompose $\mathbf{Q}$ into two matrices $\mathbf{Q}_1$ and $\mathbf{Q}_2$, then compress $\mathbf{Q}_1$ using one of the basic K-means schemes, and compress $\mathbf{Q}_2$ using the
legacy scheme, i.e., using Givens rotation and angle quantization.

\subsection{Autoencoder based schemes}\label{ae_sec}
\begin{figure*}[ht]
\begin{center}
\includegraphics[scale=0.10,width=\linewidth,trim=0cm 0cm 0cm 0cm,clip]{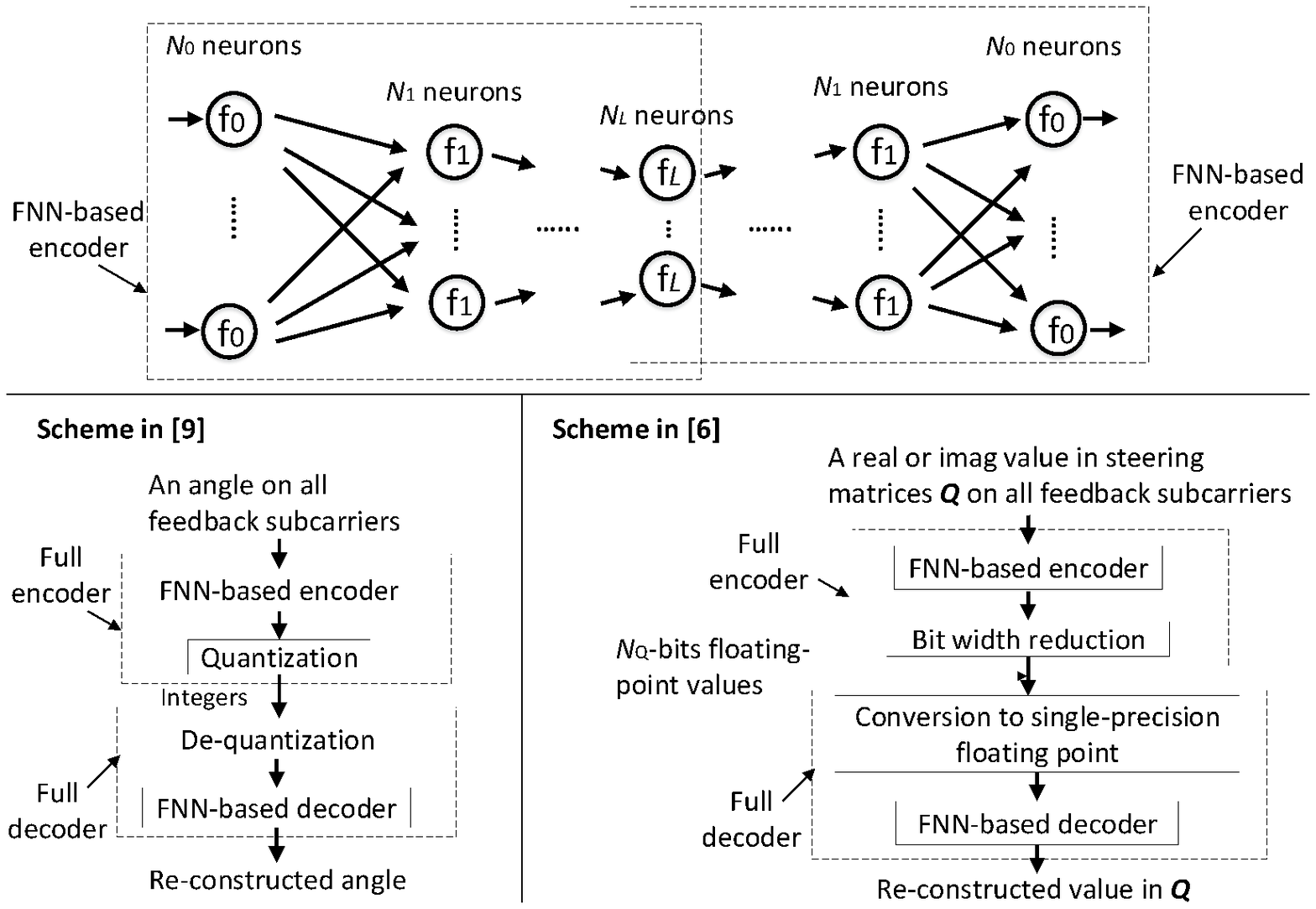}
\end{center}
\caption{A description of autoencoder based CSI compression schemes in \cite{ZHeAE2023} (or \cite{HeBFPatent2}) and \cite{SangdehAE2020}. 
The upper part of the figure describes the basic FNN-based autoencoder structure used as a component in the schemes. The lower part describes the two schemes, respectively.}
\label{ae_mthd}
\end{figure*}

Autoencoder is another unsupervised learning approach based on neural networks (NNs). 
In general, the NN training is to make adjustment of the coefficients for matching the actual output to its reference output in the labeled data. Training using labeled data is referred to as the 
the supervised learning approach. Autoencoder is a special case of NN that the reference output is the input in training, so there is no specific labeled data and thus the approach is classified as
unsupervised or self-supervised learning. 

The autoencoder based schemes in \cite{ZHeAE2023} (or \cite{HeBFPatent2}) and \cite{SangdehAE2020} are described in Fig. \ref{ae_mthd}.
The upper part of the figure describes the feedforward neural network (FNN) based autoencoder structure, which serves as a component in the two schemes. 
The lower part of the figure describes the full schemes. As depicted, the full encoder includes the FNN encoder component and a post-compression component 
(i.e., quantization in \cite{SangdehAE2020} or bit width reduction in \cite{ZHeAE2023}), 
the full decoder includes the FNN decoder component and a reverse component of the post-compression.
The FNN based autoencoder is the common part and consists of an encoder and a decoder, each of which has $(L+1)$ layers. Layer $l$ contains $N_l$ ($l=0,1,\dots,L$) neurons and each neuron has an activation function. The weighting and bias coefficients are used to connect two adjacent layers. An input and output sample of the FNN encoder is a vector with length $N_0$ and $N_L$, respectively. 
$N_0$ is the number of feedback subcarriers in the CBR and $N_L$ is the length of the compressed CSI (i.e., $N_L<N_0$).

In the training phase, the encoder output is connected with the decoder input, and the coefficients are trained/adjusted to match the encoder input to the decoder output. 
After training at the beamformer, the coefficients of the encoder are sent to the beamformee in the AIML model sharing step in Fig. \ref{aiml_mthd}.
In the two schemes, the NN coefficients are referred to as the AIML model.
In the inference phase, the beamformee uses the encoder to compress the CSI and send it in the CBR, the beamformer receives the compressed CSI and then use the decoder to uncompress the information.

The two schemes are further explained below in terms of their differences:
\begin{itemize}
\item {\bf 1) Input to the encoder}: As described in Fig. \ref{ae_mthd}, an input sample to the encoder in \cite{SangdehAE2020} is an angle $\phi$ or $\psi$ on all feedback subcarriers.
$\phi$ and $\psi$ use two separate autoecoders with different parameters (e.g., the encoder output length $N_L$), since their precision requirement is different. In the inference phase, $N_h$ samples need to be processed by the encoders representing $\phi$ and $\psi$, respectively. An input sample to the encoder in \cite{ZHeAE2023} is the real (or imaginary) value of an element in $\mathbf{Q}$. 
For example, when $N_r=8$ and $N_c=2$, the number of processed samples by the encoder is $30$. Only one autoecoder is required by the scheme in \cite{ZHeAE2023}.
\item {\bf 2) Post-compression after the FNN encoder}: Normally, the NN is performed in 32-bits single-precision floating-point format which is defined by the IEEE 754 standard.
In \cite{SangdehAE2020}, pre-defined number of bits are used to quantize the single-precision floating-point outputs to integers. 
An easy approach is to use the definition described in Sec. \ref{leg}, i.e., using $N_b$ and $(N_b+2)$ bits to quantize the encoder output for $\psi$ and $\phi$, respectively. 
The full decoder de-quantizes the received integers (bits) to single-precision floating-point values which are then processed by the FNN decoder.
In \cite{ZHeAE2023}, the bit width reduction step is to reduce the number of bits that can represent the floating-point values, i.e., from $32$ to $N_Q$ bits ($N_Q<32$).
$N_Q$-bits floating-point values are extended back to $32$ bits, before processed by the FNN decoder. 
As an example shown in \cite{ZHeAE2023}, single-precision can be reduced to half-precision ($N_Q=16$) with reasonable precision in $\mathbf{Q}$ re-construction.
To summarize, the difference between the two schemes is that the full encoder output is integer values in \cite{SangdehAE2020} but floating-point values in \cite{ZHeAE2023}.
\end{itemize}

Another scheme is proposed in \cite{GuoAE2023}, based on vector quantized variational autoencoder (VQVAE). 
Different from the above two schemes, the scheme does not assume symmetric structure at encoder and decoder, i.e., encoder and decoder use the same type of NN and even the same number of layers.
The NN type suggested in the VQVAE scheme is either convolutional NN or transformer, and the input to the encoder can be either angles or steering matrices.
In the training phase, the VQVAE at beamformer obtains a codebook and a set of NN coefficients. 
The beamformer sends the codebook as well as the NN coefficients of the encoder to the beamformee, in the AIML model sharing step in Fig. \ref{aiml_mthd}. 
In this scheme, both the codebook and the NN coefficients are referred to as the AIML model.
In the inference phase, the beamformee uses the codebook and the encoder to generate the codeword indices, and then the beamformer uses the indices, the codebook and the decoder to reconstruct $\mathbf{Q}$. 

\subsection{AIML model sharing}\label{mdl_sharing}
As discussed in Sec. \ref{kmeans_sec} and \ref{ae_sec}, the AIML model sharing is referred to as the additional
communication signaling overhead required to send the codebook or/and NN coefficients from the beamformer to the beamformee.
Essentially, the AIML CSI compression schemes can be seen as methods to transfer a large CSI feedback overhead to a smaller feedback overhead and the model sharing overhead.
The model sharing overhead affects the goodput and need to be carefully considered in the system design.

In Sec. \ref{kmeans_sec}, a simple codebook compression method has been discussed for the K-means schemes.
This subsection focus on the compression of NN coefficients for the autoencoder schemes.
NN quantization is an approach to compress NN and has been accepted for consideration in the AIML model sharing use case, to reduce the signaling overhead \cite{XWangAIMLRprt2023}.
It is used to reduce the number of bits that representing the coefficients (i.e., from $32$ to a smaller value), and includes two methods: 
the post-training quantization (PTQ) and the quantization-aware training (QAT).
PTQ performs quantization after training is complete, thus it is very effective and fast to implement because it does not require retraining of the NN. 
However, it may not be enough to mitigate the large quantization error introduced by low-bit quantization. 
QAT performs quantization together with the training and is aiming for low-bit quantization, such as $4$-bits and below.
In \cite{ZHeNNQ2023}, an example is given to apply the NN quantization to the autoencoder scheme in \cite{ZHeAE2023}.
It is shown that by applying $8$-bits PTQ to the NN coefficients of the encoder, the communication overhead in the AIML model sharing is reduced by $74$\% 
without degrading the re-construction precision of the steering matrix.

\subsection{Key performance indicators}\label{kpi}
There are four key performance indicators (KPIs) defined by the CSI compression use case in \cite{XWangAIMLRprt2023}:
$1$) CSI feedback overhead; $2$) additional overhead introduced by the AIML schemes (e.g., model sharing in Sec. \ref{mdl_sharing});
$3$) PER performance for the beamformed data transmission; $4$) computational complexity of CSI compression.
More complicated than the legacy scheme described in Sec. \ref{leg}, the design logic of the AIML schemes 
is to find a good trade-off among feedback overhead, model sharing overhead, PER performance and even the compression complexity.

\begin{figure*}[ht]
\begin{center}
\includegraphics[scale=0.12,width=\linewidth,trim=0cm 0cm 0cm 0cm,clip]{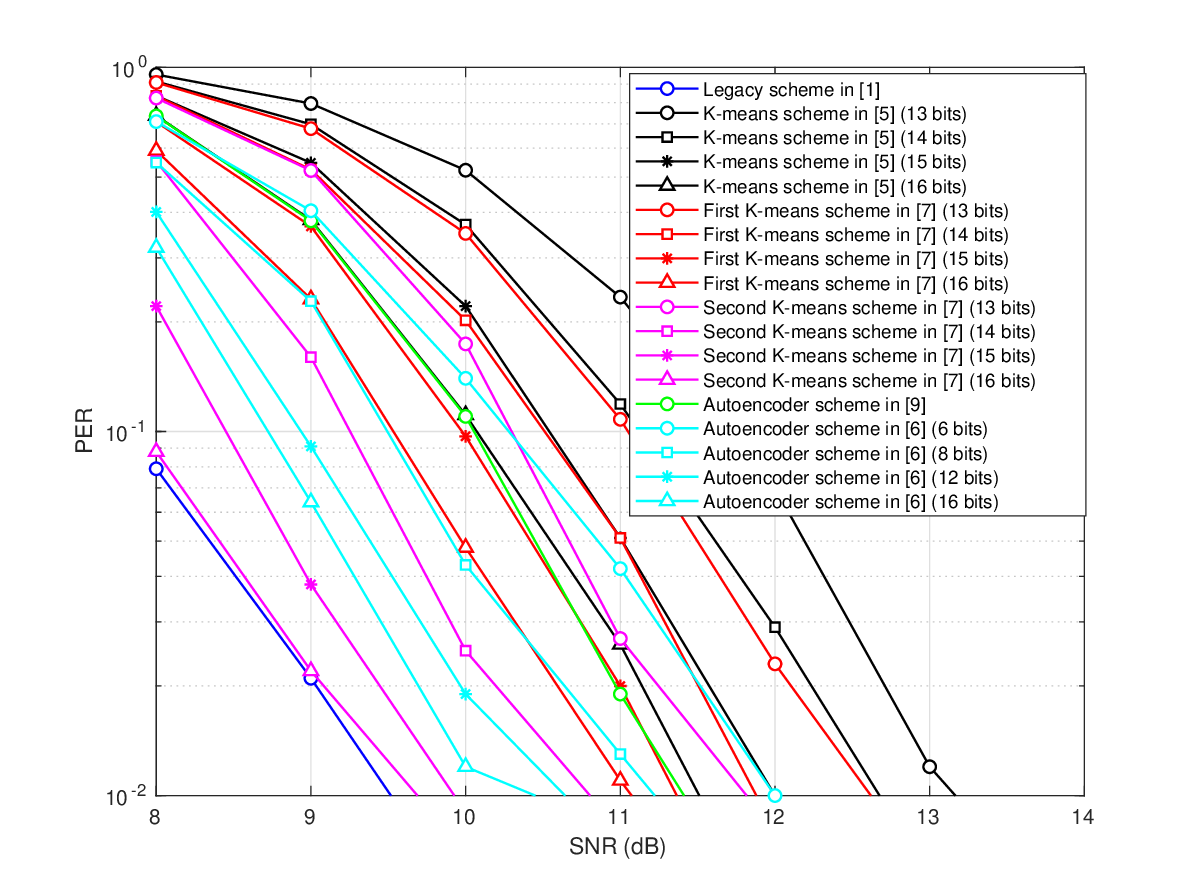}
\end{center}
\caption{PER performance of various CSI compression schemes as a function of the mean SNR (the number of bits in the legend is $N_{bf}$ and $N_{L}$ for the K-means and autoencoder schemes,
respectively)}
\label{per_res_ieee}
\end{figure*}

The PER performance as a function of the SNR is shown in Fig. \ref{per_res_ieee}, for various CSI compression schemes.
In the simulation results, the common evaluation conditions used for all schemes are: $20$ MHz 11ax signal, modulation and coding scheme (MCS) $3$, $N_r=8$, $N_c=2$, 802.11 channel model D,
$64$ feedback subcarriers per CBR, $1000$ bits packet payload size in the beamformed data transmission.
The number of bits in the legend is $N_{bf}$ and $N_{L}$ for the K-means and autoencoder schemes, respectively.
For the K-means schemes, $N_b=4$ is used for the codebook compression, for the autoencoder schemes, NN quantization mentioned in Sec. \ref{mdl_sharing} is not used.
For the autoencoder scheme in \cite{SangdehAE2020}, design parameter $N_L=32$ and $16$ are used for $\phi$ and $\psi$, respectively.
For the autoencoder scheme in \cite{ZHeNNQ2023}, half-precision floating-point format is used ($N_Q=16$).
For the legacy scheme and the autoencoder scheme in \cite{SangdehAE2020}, $N_b=4$ is used for feedback compression.
Moreover, the communication overhead and CSI compression complexity for each curve in Fig. \ref{per_res_ieee} is summarized in Tab. \ref{ovhd_cmplx_tab}.
The complexity is evaluated in terms of the number of required real-value multiplications in the CSI compression per CBR, since multiplication is the most complex operation \cite{He2022}.
The observations and comments are summarized below:
\begin{itemize}
\item In terms of PER performance, all AIML enabled compression schemes are worse than the legacy schemes. However, the key benefit of using AIML scheme is to reduce the CSI feedback overhead.
\item For the K-means schemes, the second scheme in \cite{ZHeKmeans2023} has the best PER performance, 
and the first scheme in \cite{ZHeKmeans2023} is slightly better than the scheme in \cite{ZLinKeans2022}. The first K-means scheme in \cite{ZHeKmeans2023} has the largest feedback overhead but the smallest AIML model sharing overhead, the other two schemes have the same overhead. The second scheme in \cite{ZHeKmeans2023} has the highest complexity and the other two have the same complexity.
\item For the autoencoder schemes, the scheme in \cite{ZHeAE2023} is better than the scheme in \cite{SangdehAE2020} in terms of PER, model sharing overhead and complexity. However, its CSI feedback overhead is larger.
\item For the AIML schemes, compress steering matrices directly (i.e., second K-means scheme in \cite{ZHeKmeans2023} and the autoencoder scheme in \cite{ZHeAE2023}) 
provides better PER performance than compressing the angles. 
Compress steering matrices directly avoids the angles computation from the steering matrices, thus potentially provides a lower complexity.
\item In the AIML schemes, the autoencoder scheme in \cite{ZHeAE2023} has the lowest model sharing overhead and complexity but the largest feedback overhead. 
It is the only AIML scheme with a complexity lower than that of the legacy scheme, and thus it can be used to address {\bf Challenge II} in Sec. \ref{intro_sec}.
\end{itemize}

Since the standardization for AIML enabled CSI compression is still at its early stage, the goodput performance is not given here. 
At this stage, it is very difficult to evaluate the goodput without a standardized higher layer signaling procedure for the AIML model sharing and the AIML enabled CBR. 
For examples, what the packet structure should be used to send the model and the CBR, what MCS should be used for these overheads, how often the model and CBR packets should be sent etc. 
However, there are limited goodput results in the literature with assumptions:
In \cite{ZLinKeans2022}, the goodput is evaluated without considering the model sharing overhead and a fixed low MCS for CBR (i.e., large CBR overhead), 
it is claimed that the goodput has been improved by up to $52$\%. In \cite{ZHeKmeans2023}, a more realistic MCS is used for CBR, it is found that the goodput has been improved by up to $22$\%.
However, a realistic AIML model sharing procedure and the CBR frequency are not well addressed in the literature.
\begin{table*}[ht]
  \centering
  \caption{A comparisons of communication overhead and CSI compression complexity}
  \label{ovhd_cmplx_tab}
  \begin{tabular}{p{3cm}p{3cm}p{3cm}p{3cm}}
    \hline
    \textbf{Schemes} & \textbf{Number of CSI feedback bits per CBR} & \textbf{Number of bits per AIML model sharing} & \textbf{Number of multiplications per CBR} \\
    \hline
    Legacy scheme in \cite{wifi6} & $8320$ & $0$ & $225$K \\
    \hline
    K-means scheme          & $832$ ($N_{bf}=13$) & $1065$K & $13857$K \\
    in \cite{ZLinKeans2022} & $896$ ($N_{bf}=14$) & $2130$K & $27488$K \\ 
                            & $960$ ($N_{bf}=15$) & $4260$K & $54751$K \\
                            & $1024$ ($N_{bf}=16$) & $8520$K & $109277$K \\
    \hline
    First K-means scheme    & $1664$ ($N_{bf}=13$) & $639$K & $13857$K \\
    in \cite{ZHeKmeans2023} & $1792$ ($N_{bf}=14$) & $1278$K & $27488$K \\
                            & $1920$ ($N_{bf}=15$) & $2556$K & $54751$K \\
                            & $2048$ ($N_{bf}=16$) & $5112$K & $109277$K \\
    \hline
    Second K-means scheme   & $832$ ($N_{bf}=13$) & $1065$K & $15729$K \\
    in \cite{ZHeKmeans2023} & $896$ ($N_{bf}=14$) & $2130$K & $31457$K \\
                            & $960$ ($N_{bf}=15$) & $4260$K & $62914$K \\
   		           & $1024$ ($N_{bf}=16$) & $8520$K & $125829$K \\
    \hline
    Autoencoder scheme     & $3328$ & $879$K & $578$K \\
    in \cite{SangdehAE2020} & ($N_L=32$ for $\phi$ &  & \\
                            & $N_L=16$ for $\psi$) &  & \\
    \hline
    Autoencoder scheme  & $2880$ ($N_L=6$)  & $91$K & $84$K  \\
    in \cite{ZHeAE2023} & $3840$ ($N_L=8$)  & $94$K & $86$K  \\
                        & $5760$ ($N_L=12$) & $99$K  & $91$K  \\
                        & $7680$ ($N_L=16$) & $104$K  & $96$K  \\
    \hline
  \end{tabular}
\end{table*}

\section{Future research directions}\label{future_sec}
In the procedure of 802.11 standardization, TIG is the earliest stage, the next step is formation of a SG, whose purpose is to develop a Project Authorization Request (PAR) that set out the scope of a proposed amendment to the 802.11 standard. Once this is approved, a TG is formed, whose purpose is to develop the amendment, in line with the PAR.
The current focus of AIML TIG is to define the use case and its KPIs at a higher level, there are more detailed technical works need to be done, e.g., in the SG stage.
To encourage more researchers/engineers working in this area, a list of future research directions are summarized below:
\begin{itemize}
\item As mentioned in Sec. \ref{kpi}, a realistic evaluation method for goodput need to be established, so that all the schemes can be compared fairly in terms of goodput and complexity.
\item The evaluation of the feedback latency introduced by the compression complexity has not been discussed in the literature. 
It is difficult to evaluate this using simulations, a real-time testbed can be developed for this purpose.
\item All the existing results in the literature use 11ax as the baseline. When 11be is frozen, it is better to use it as the baseline, so that a larger transmit antennas ($N_r>8$) can be evaluated. 
\item All the existing results consider single-user multiple input multiple output (MIMO) case, it is interesting to see the performance in multi-user MIMO case where more CSI feedback overhead is used.
\item The compression complexity of the K-means schemes are high, this is because Euclidean distance is used as the searching metric. A new searching metric could be proposed to reduce the complexity.
If such complexity can be significantly reduced, the second scheme in \cite{ZHeKmeans2023} may have lower complexity than that of the legacy scheme, 
since it does not require angle computation from the steering matrices.  
\item QAT in NN quantization can be considered for the autoencoder schemes to reduce model sharing overhead. 
\end{itemize}

\section{Conclusions}\label{concl_sec}
In this paper, we have given an overview of the current standardization activities regarding the AIML enabled beamforming CSI compression technique, defined by the IEEE 802.11 AIML TIG.
Firstly, we have introduced the background of the legacy beamforming in the existing WiFi standards, and its technical challenges the AIML TIG is going to tackle in future WiFi generations.
Secondly, we have provided more details on the legacy CSI compression scheme followed by a couple of AIML enabled compression schemes defined by the TIG. 
Moreover, we have listed the defined KPIs for the technique, and use simulated and analyzed data to explain how the AIML schemes are designed according to the KPIs.
Finally, we have given some future research directions which have not been addressed by the TIG.
We have encouraged more researchers and engineers to contribute to the next generations of WiFi beamforming technique.

\bibliography{mybib}
\bibliographystyle{ieeepes}

\begin{IEEEbiographynophoto}{Ziming He (Senior Member, IEEE)}
received the B.Eng. degree from Wuhan University, Wuhan, China, in 2006, and the Ph.D. degree from the University of Surrey, Guildford, U.K., in 2012. He is currently a Senior Staff System Engineer with the Samsung Cambridge Solution Centre, System LSI, Samsung Electronics, Cambridge, U.K., and has developed many cutting-edge algorithms for Samsung’s Exynos modem chipset products. He is the inventor of more than ten granted/pending U.S. patents, a Voting Member of the IEEE 802.11 Standard Association working groups, and a member of the Bluetooth Special Interest Group. He is extensively involved in the U.K., founded research projects, and has authored/coauthored more than 40 academic articles. He is a fellow of IET and the U.K. Chartered Engineer.
\end{IEEEbiographynophoto}

\end{document}